# A Fully Tunable Single-Walled Carbon Nanotube Diode

*Chang-Hua Liu, Chung-Chiang Wu, and Zhaohui Zhong\**

Department of Electrical Engineering and Computer Science, University of Michigan, Ann Arbor, Michigan 48109

\*Corresponding author. Electronic mail: zzhong@umich.edu

ABSTRACT

We demonstrate a fully tunable diode structure utilizing a fully suspended single-walled carbon nanotube (SWNT). The diode's turn-on voltage under forward bias can be continuously tuned up to 4.3 V by controlling gate voltages, which is ~6 times the nanotube bandgap energy. Furthermore, the same device design can be configured into a backward diode by tuning the band-to-band tunneling current with gate voltages. A nanotube backward diode is demonstrated for the first time with nonlinearity exceeding the ideal diode. These results suggest that a tunable nanotube diode can be a unique building block for developing next generation programmable nanoelectronic logic and integrated circuits.

Integrated nanoelectronics [1] of the coming generation can benefit from actively tunable device components, where a single device can be programmed to embody different device concepts to achieve high packing density, diverse functionality, and low power consumption. However, electronic characteristics of conventional devices are pre-determined by material properties such as bandgap energy, doping level, metal-semiconductor interfaces, etc, and active tuning of individual device remains extremely challenging. For instance, a conventional diode has fixed rectification characteristics; the intrinsic doping levels across the junction determine the built-in potential and the turn-on voltage of the diode. To active modulate junction properties, including doping concentrations and doping profile in bulk devices, are limited mainly by the strong screening effect and the increasing 3-dimnesional density-of-states (DOS) with increasing energy. To this regard, recent works address these issues by chemical or electrochemical doping and demonstrate the concept of tunable electronics [2-4]. But these device structures and fabrication methodologies are difficult to be integrated with existing CMOS platform.

SWNTs are quasi one-dimensional structures with promising electrical and optical properties [5-7]. Because the DOS decreases with increasing energy in one-dimension, doping concentration in carbon nanotube can be easily modulated through electrostatic gating [8-9]. Ideal diode behavior has been demonstrated on suspended SWNTs with electrostatic p-n doping [10-12], and efficient light emission and photocurrent generation have also been shown in SWNT diodes [13-15]. In addition, strong band-to-band tunneling across nanotube Schottky junction or p-n junction [16-18] has lead to SWNT tunneling diodes exhibiting negative differential resistance with high peak to valley ratio [19]. However, all previous nanotube diode design can only exhibit one specific diode concept among diverse types of junction diodes [20].

In this work, we demonstrate a fully tunable SWNT diode concept, which incorporates several types of junction diodes within single device architecture by simply controlling gate voltages. Figure 1a illustrates the schematic of our fully suspended SWNT p-n diode device. To fabricate this device, we adopt a unique one-step direct transfer technique [12]. Briefly, a pristine SWNT synthesized by chemical vapor deposition is directly transferred from the growth substrate to the device substrate with pre-patterned cathode, anode, and two bottom split gate electrodes. The electrostatic doping of the nanotube is achieved by controlling the two bottom split gate voltages, $V_{g1}$ and $V_{g2}$. Figure 1b shows a scanning electron microscopy (SEM) image of a fully suspended SWNT diode. Our previous work has demonstrated that the ideality factor of the fully suspended SWNT diodes can reach 1 [12], attributed to the elimination of surface induced states and the adoption of pristine SWNTs [10]. Here, we focus on further investigation of diode's tunability, and data measured at room temperature and under ambient condition are presented.

Figure 1c shows *I-V* characteristics of a representative SWNT diode with split gate separation of 4 μm under different gate bias conditions. Positive $V_{g1}$ and negative $V_{g2}$ electrostatically dope nanotube sections above the split gates into n and p type, respectively. The formation of the p-n junction leads to the rectified *I-V* curves which turn on at forward bias direction. Interestingly, when holding $V_{g2}$ at a constant voltage of -4 V and increasing $V_{g1}$ from 2 to 9 V, the rectified *I-V* curves shift gradually toward higher turn-on voltages while maintaining the same nonlinearity. The extracted turn-on voltages show linear dependence on the split gate voltage $V_{g2}$ (Fig. 1d), with maximum value exceeding 4.3 V, corresponding to ~6 times the bandgap voltage of the nanotube. We also studied the diode characteristics under constant $V_{g1}$ of 4 V and varying $V_{g2}$ from -2 to -9 V (Fig. 1e). As $V_{g2}$ decreases to more negative voltages, the turn-on voltages drop from 2.6 V to 0.5 V. The extracted turn-on voltages again show linear dependence on the amplitude of split gate voltage, $-V_{g2}$, but with a negative slope (Fig. 1f). Importantly, the observed gate-controlled tunability and the beyond-bandgap voltage rectification are not possible in conventional diodes.

In order to understand the unique adjustable turn-on voltages in SWNT diode, we first evaluate the possibility of a tunable and above-bandgap built-in junction potential due to electrostatic gating. For a typical p-n diode, the built-in potential is determined by the Fermi-level difference across the junction, which in our case can be tuned by the split gate voltages. As a result, a greater potential difference between $V_{g1}$ and $V_{g2}$ will induce a higher built-in potential barrier and hence a larger diode turn-on voltage. This prediction agrees with the results shown in the Figure 1d, but opposes to the results shown in the Figure 1f. Furthermore, the turn-on voltage of the nanotube diode can exceed 4 V. To achieve this high built-in potential, the nanotube would have been degenerate doped into third subband in both n and p sections. However, with the small gate capacitance of our fully suspended device, ~ $7\times10^{-18}$ Farad/μm estimated by the geometry, Fermi-level can only be shifted to the first subband edge under 10V gate voltage. Therefore, the nanotube diode turn-on voltage cannot be simply explained by the junction built-in potential as in a conventional semiconductor diode.

In order to understand the *I-V* characteristics of our devices, it is necessary to examine the injected carriers from the complete band diagram including metal/nanotube contacts. For a fully suspended nanotube diode, p-n junction can be formed by electrostatic doping using the split gates, and in addition, Schottky junction will be also formed at the metal/nanotube interfaces. Using high work function metal Au as the contact, there will be a Schottky barrier for electron injection while hole injection remains Ohmic. Previous studies have shown that Schottky barrier width of the metal/SWNT interface can be significantly reduced when the nanotube is coupled with the gate electrode through thin high-$\kappa$ dielectrics [17, 21-22]. Thus, thermally assisted tunneling for both electrons and holes can be enhanced for thin enough barrier. However, strong electrons tunneling through Schottky barrier is less likely for our fully suspended nanotube devices with small gate capacitance and weak gate coupling. As the consequence, holes injected from the anode are the dominating carriers for current flow under forward bias.

The energy band diagram under equilibrium is illustrated in Figure 2a (black). For a conventional diode with Ohmic contact, forward bias voltage reduces the potential barrier for carrier injection across the junction, and leads to exponential increase in diode current. In our device, however, forward bias voltage first drops across the n-side Schottky junction instead of the p-n junction. The diode current is determined by the recombination of injected holes with electrons in n-region. When the effective length of n-region $l_n$ is much greater than the hole minority carrier diffusion length $L_p$, the diode current will remain more or less constant. Further increase in forward bias will reduce $l_n$, and when $l_n$ approaches $L_p$, $l_n \sim L_p$, the injected holes can be swapped across the Schottky junction under large field, and the diode will start turning on (Fig. 2b). To confirm this prediction, we plot diode $I$-$V$ curves in log scale, and as shown in Fig. 2c, the diode current indeed remains roughly constant until forward bias reaching the turn on threshold voltage. Furthermore, $l_n$ can also be modulated by the effect of fringing field from $V_{g1}$ and $V_{g2}$. More positive $V_{g1}$ increases the electron doping concentration, and at the same time increases $l_n$ (Fig. 2a, blue), leading to a larger diode turn-on voltage. On the other hand, more negative $V_{g2}$ increases the hole doping concentration but reduces $l_n$ (Fig. 2a, yellow), leading to a smaller turn-on voltage. Our experimental results presented in Figure 1c-1f once again agrees with the prediction. These results confirm that the turn-on voltage tuning in the fully suspended SWNT diodes is achieved by modulating hole extraction in n-doped region.

We also explored the tunability of reverse bias characteristics for the fully suspended SWNT diode. For conventional p-n diodes, tunneling current is usually suppressed due to the width of space charge region, which is made out of immobile donor and acceptor ions. In contrast, for electrostatically doped nanotube p-n diodes, all charges are mobile carriers and they tend to accumulate at the interface. Therefore, by bringing two split gates closer and/or applying higher gate voltages to create a sharper p-n junction, it is possible to strongly enhance the band-to-band tunneling current [18, 23].

Figure 3a shows the *I-V* curves for another fully suspended SWNT diode with split gate separation of 1 μm. $V_{g2}$ is held at constant potential of -3 V, while $V_{g1}$ is increased from 6 V to 10 V. In the forward bias region, this diode exhibits similar tunability as the device shown earlier. However, the reverse bias leakage current is strongly enhanced by 50 times compared to previous device with 4 μm split gate separation. Moreover, leakage current increases with increasing gate voltage of $V_{g1}$. These results agree with enhanced band-to-band tunneling across a sharper p-n junction under reverse bias. More careful examinations of reverse bias *I-Vs* reveal that the tunneling current increases exponentially from 0 V to -0.2 V, and then grows linearly at higher reverse bias voltage (Fig. 3a, inset).

To explain this unusual *I-V* relation, we use Fermi's golden rule to model the band-to-band tunneling current [20]: $I_{tunneling} \alpha \int_{E_{cn}}^{E_{vp}} (F_c(E) - F_v(E)) T N_c(E) N_v(E) dE$. Here, *F$_c$(E)* and *F$_v$(E)* are Fermi-Dirac distribution functions at room temperature; T is the tunneling probability through the junction potential barrier; *N$_c$(E)* and *N$_v$(E)* are density-of-states (DOS) of carbon nanotube for the conduction band and valence band, respectively. The equation integrates from the conduction band edge in the n side ($E_{cn}$) to the valence band edge in the p side ($E_{vp}$), and the tunneling current is decided by the overlap DOS ($\int N_c(E)N_v(E)$), modulated by Fermi energy difference ($\int F_c(E) - F_v(E)$) between p and n side. Only interband tunneling within the first subband is taken into consideration, and tunneling within higher subbands is negligible due to relatively low tunneling probability [18]. The simulation result is shown in the Fig. 3b, which reveals the same *I-V* characteristics as our experimental results, with an exponential current increase followed by linear dependence on reverse bias voltage.

The I-V relation can also be qualitatively understood from the changes in overlap DOS of conduction band and the valance band (Fig 3b, inset). The overlap DOS has maximum value at the band edge due to Van Hove singularity. Under small reverse bias voltage, Fermi-level of p side ($E_{Fp}$) shifts across the region with high overlap DOS. Thus, the tunneling junction is highly conducting. As reverse bias increases, $E_{Fp}$ shifts into the low overlap DOS region, causing tunneling current increases linearly with increasing bias. We note that these unusual band-to-band tunneling features are unique to one dimensional junction.

Another peculiar consequence arising from the enhanced band-to-band tunneling is that, the reverse bias current can be much greater than forward bias current within small voltage range. One representative *I-V* curve is shown in Figure 4 with $V_{g1}$ = 6 V and $V_{g2}$ = -6 V. The forward bias current is suppressed below 10 pA with $V_{bias}$ up to 1 V, the consequence of minimized hole extraction. Interestingly, the reveres bias current increases exponentially to 150 pA at $V_{bias}$ = -0.23 V, which is 30 times greater than the forward bias current. The results are in strong reminiscence of a backward diode [20], which rectifies under forward bias but conducts under reverse bias. It is notable that our SWNT backward diode has large turn-on voltage (~1 V) and small leakage current (~5 pA) in the forward bias region, both of which are desirable for backward rectification. Most importantly, fast increasing tunneling current in the reverse bias region leads to large zero-bias nonlinearity, which is especially challenging to achieve for conventional semiconductor backward diodes [20]. We further calculate the nonlinear curvature coefficient, $\gamma$, which is defined as $\gamma = (\partial^2 I / \partial V^2)/(\partial I / \partial V)$ (Fig. 4, inset). At zero bias voltage, the nanotube backward diode exhibits a curvature coefficient of 34.3 $V^{-1}$, approaching the thermionic limit value of $\gamma = q/KT$ = 38.5 $V^{-1}$ at 300 K. At $V_{bias}$ ~ 15 mV, $\gamma$ can even go beyond thermionic limit and reach maximum value of 46.4 $V^{-1}$. These results suggest remarkable potential for SWNT backward diode's applications in high speed switching, microwave mixing and detection, and small amplitude rectification [24, 25]. Combined with its full tunability in both forward and reverse bias directions, the programmable SWNT diodes should open up new opportunities for integrated nanoelectronics and nanophotonics.

The authors gratefully acknowledge Prof. John Hart and Dr. Yongyi Zhang for assistance in carbon nanotube synthesis. We thank the support from the U-M/SJTU Collaborative Research Program in Renewable Energy Science and Technology. The devices were fabricated at the Lurie Nanofabrication Facility at the University of Michigan, a member of the National Nanotechnology Infrastructure Network funded by the National Science Foundation.

[22]  J. Appenzeller *et al.*, Phys. Rev. Lett. **89**, 126801 (2002).

[23]  J. Knoch, and J. Appenzeller, Physica Status Solidi A **205**, 679 (2008).

[24]  J. N. Schulman, and D. H. Chow, IEEE Elec. Dev. Lett. **21**, 353 (2000).

[25]  S. Y. Park *et al.*, Electronics Lett. **43**, 295 (2007).

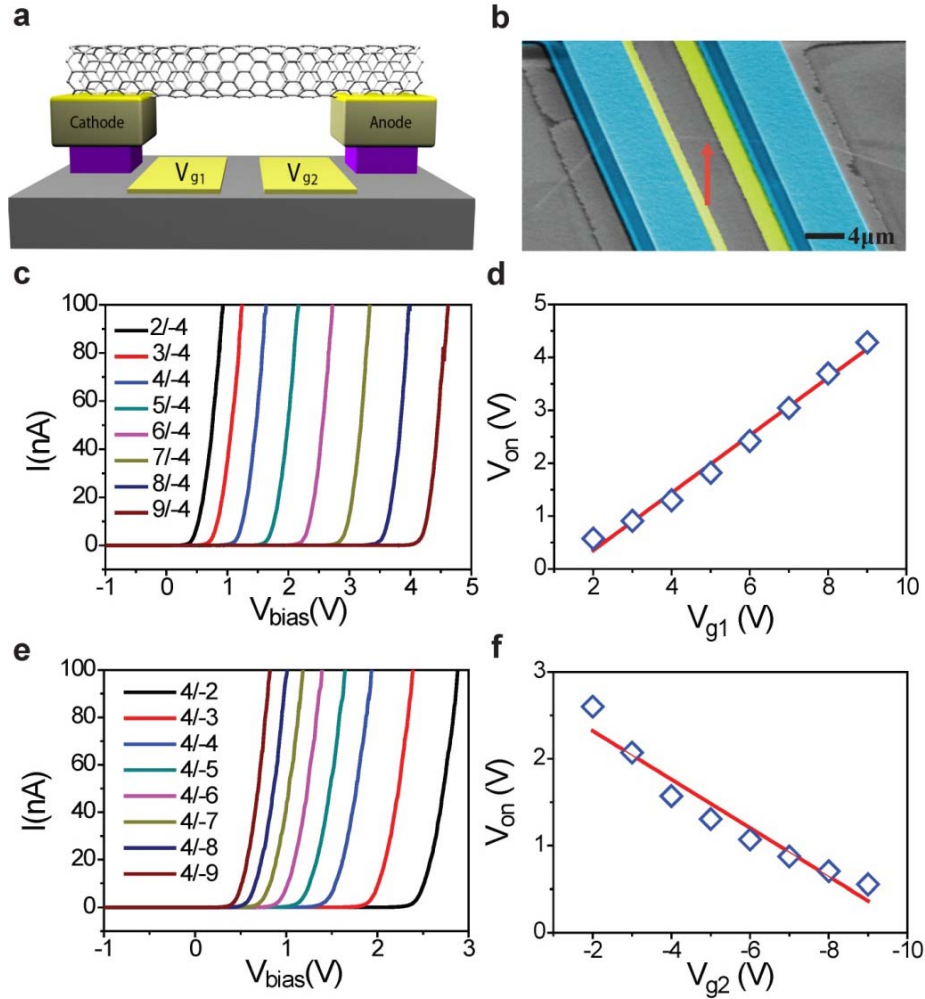

FIG. 1 (color online). (a) Schematic drawing of a fully suspended SWNT diode. The nanotube is suspended between cathode and anode electrodes, and located 1.5 μm above the bottom gate electrodes. 30-nm-thick gold is used as a nanotube contact metal. A pair of split gates $V_{g1}$ and $V_{g2}$ are separated by 4 um, and used to electrostatically dope SWNT. (b) SEM image (false colored) of the suspended SWNT device. The arrow indicates the position of the nanotube. Blue regions correspond to cathode and anode electrodes, and yellow regions correspond to bottom gates. (c) *I-V* characteristics of the SWNT diode measured for different $V_{g1}$ gate bias voltages. $V_{g2}$ is biased at constant voltage of -4 V, with $V_{g1}$ voltage increasing from 2 V to 9 V. (d) The dependence of turn-on voltages on $V_{g1}$ gate voltages. (e) *I-V* characteristics of the SWNT diode measured for different $V_{g2}$ gate bias voltage. $V_{g1}$ is biased at constant voltage of 4 V, with $V_{g2}$ voltage decreasing from -2 V to -9 V. (f) The dependence of turn-on voltages on $V_{g2}$ gate voltages.

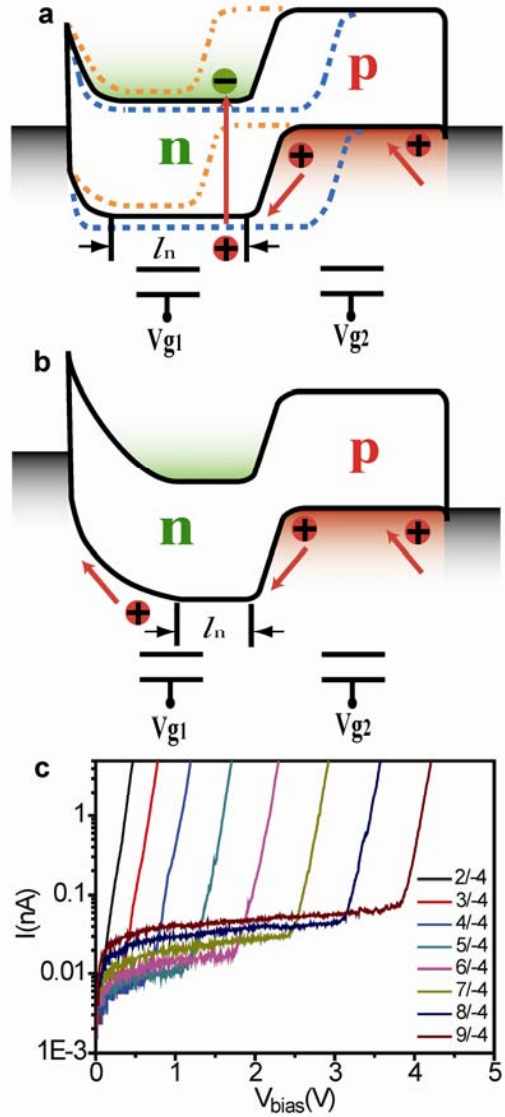

FIG. 2 (color online). (a) Energy band diagram of the SWNT diode under equilibrium (black line). $l_n$ denotes the n channel length. $l_n$ increases with higher $V_{g1}$ gate voltage (blue line), and decreases with lower $V_{g2}$ gate voltage (yellow line). When $l_n$ is much greater than the hole minority carriers diffusion length $L_p$, hole injected across the junction is recombined with electron. (b) Energy band diagram under forward bias voltage. The external bias voltage causes band bending in the n section and reduces $l_n$. Once $l_n$ becomes comparable to $L_p$, injected hole will be swapped to the left contact under electric field, and the p–n junction will be turned on. (c) Log scale plot of the same *I-V* curves shown in Fig. 1(c).

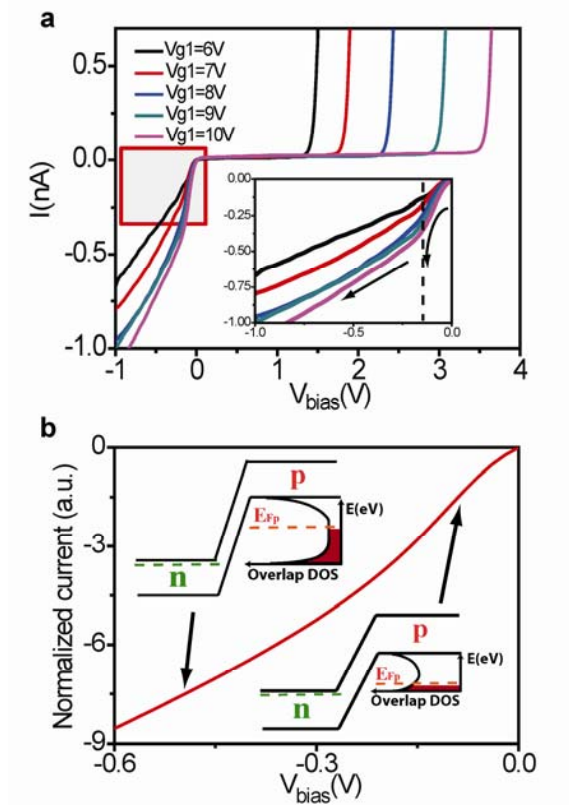

FIG. 3.

FIG. 3 (color online). (a) *I-V* characteristics of a SWNT diode with two bottom gates separated by 1μm. The device shows similar tunability in the forward bias region as in previous device, and also a strong tunneling current in the reverse bias region. The inset shows the zoom-in view of the reverse bias region. (b) Simulated band-to-band tunneling current for the one dimensional SWNT diode. The inset shows the overlap DOS under small and large reverse bias voltages, respectively.

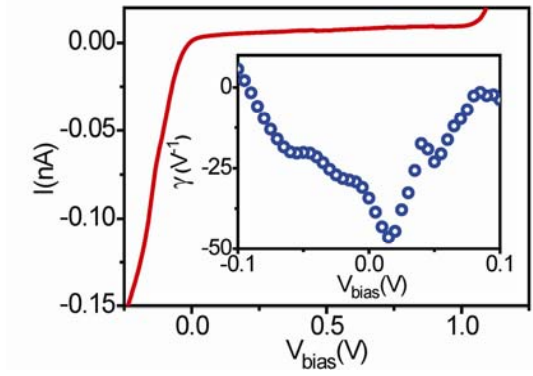

FIG. 4 (color online). By applying proper gate voltages, the *I-V* characteristics of the SWNT diode show backward rectification behavior. The inset shows the measured curvature coefficient *γ* versus bias voltage. The maximum curvature coefficient exceeds the theoretical value for an ideal diode.